\documentclass[aps,prd,reprint,amsmath,amssymb,superscriptaddress,floatfix]{revtex4}
\usepackage{graphicx}
\usepackage{verbatim} 
\usepackage{amsfonts}
\usepackage{amssymb}
\usepackage{rotating}
\usepackage{booktabs}
\usepackage{xcolor}
\usepackage{soul}
\usepackage{color}
\usepackage{slashed}
\usepackage{multirow}
\usepackage{makecell}
\usepackage{epsf}
\usepackage{ulem}
\usepackage{cancel}
\usepackage{color,bm}
\usepackage[colorinlistoftodos]{todonotes}
\usepackage{diagbox}
\usepackage[colorlinks=true,citecolor=cyan,urlcolor=blue,bookmarks=true,bookmarks=true,bookmarksopen=true,bookmarksnumbered=true,bookmarksopenlevel=3]{hyperref}
\usepackage{subfigure}
\definecolor{airforceblue}{rgb}{0.36, 0.54, 0.66}
\definecolor{steelblue}{rgb}{0.27, 0.51, 0.71}
\definecolor{amber}{rgb}{1.0, 0.49, 0.0}

\begin{document}

\title{$Z'$ Mediated right-handed Neutrinos from Meson Decays at the FASER}

\author{Jiale Li}
\affiliation{Institute of Theoretical Physics, School of Physics, Dalian University of Technology, No.2 Linggong Road, Dalian, Liaoning, 116024, P.R.China }

\author{Wei Liu}
\email{corresponding author, wei.liu@njust.edu.cn}
\affiliation{Department of Applied Physics and MIIT Key Laboratory of Semiconductor Microstructure and Quantum Sensing, Nanjing University of Science and Technology, Nanjing 210094, China }

\author{Hao Sun}
\email{corresponding author, haosun@dlut.edu.cn}
\affiliation{Institute of Theoretical Physics, School of Physics, Dalian University of Technology, No.2 Linggong Road, Dalian, Liaoning, 116024, P.R.China }

\date{\today}

\begin{abstract}
We investigate the pair production of right-handed neutrinos mediated by a $Z^\prime$ mainly from the meson decays at the FASER detector of the HL-LHC. The $Z^\prime$ can be the additional gauge boson in either the $U(1)_{B-L}$ or sterile $\nu$-specific $U(1)_s$ model. Taking the gauge coupling or the kinetic mixing at the current limits, we analyse the sensitivity to the masses of the heavy neutrinos, $M_N$, and active-sterile mixing, $|V_{lN}|^2$, of the FASER2. In a background free scenario, for the $U(1)_{B-L}$ case, FASER2 is able to probe $|V_{lN}|^2 \approx 10^{-8}$ when $M_N \sim 0.3$ GeV, which is comparable to the current limits from the beam dump experiments. When comes to the $U(1)_s$ model, FASER2 can  probe $|V_{lN}|^2 \approx 10^{-10}$, which is better than the current limits for at least one magnitude, in all three flavours.
A proposed long-lived particle detector, FACET, is also studied, while no significant difference from FASER2 is derived.

\vspace{0.5cm}
\end{abstract}
\maketitle
\setcounter{footnote}{0}

\section{Introduction}
In the Standard Model~(SM), neutrinos are predicted to be massless fermions. Nevertheless, tiny neutrino masses have been observed at neutrino oscillation experiments, pointing the existence of the physics beyond the SM~\cite{Super-Kamiokande:1998kpq, SNO:2002tuh}. To explain the mass of neutrino, the seesaw mechanisms were proposed, arguing that neutrinos masses are produced by adding massive right-handed neutrinos~(RHN). In this way Dirac and Majorana mass terms can be added, thus giving the light neutrino mass with natural Yukawa couplings.

The RHNs can mix to the active neutrinos via the mixing $V_{lN}$, and $m_\nu\sim{V_{lN}}^{2}M_N$ according to the canonical type-I seesaw. Various theoretical and experimental studies have been done to search for the RHNs~\cite{Batell:2016zod, Bhattacherjee:2021rml, Accomando:2017qcs, Das:2019fee, Cheung:2021utb, Chiang:2019ajm, FileviezPerez:2020cgn, Das:2018tbd, Maiezza:2015lza, Nemevsek:2016enw, Mason:2019okp, Accomando:2016rpc, Gao:2019tio, Gago:2015vma, Jones-Perez:2019plk, Liu:2021akf, Deppisch:2018eth, Amrith:2018yfb, Deppisch:2019kvs, Liu:2022kid,Liu:2022ugx, Zhang:2023nxy, Liu:2023nxi, Das:2017nvm, Bhattacherjee:2023plj, Barducci:2023hzo}.
From the existing experiments, including the ones from beam-dump and collider experiments, the current limits set $|V_{lN}|^2 \lesssim 10^{-8}$ for RHNs with masses of $\mathcal{O}(0.1)$~GeV~\cite{Bolton:2019pcu}. In such case, the RHNs can be long-lived particles~(LLP), hence can lead to unique signatures at the LHC, with final states produced meters away from the interaction points.

Several detectors at the LHC, aiming for detecting such LLPs have been proposed, including the FASER~\cite{Feng:2017uoz}, FACET~\cite{Cerci:2021nlb}, MATHUSLA~\cite{Chou:2016lxi}, MoEDAL-MAPP~\cite{Frank:2019pgk}, CODEX-b~\cite{Gligorov:2017nwh}, ANUBIS~\cite{Hirsch:2020klk} and etc. Among them, FASER and MoEDAL-MAPP are already operating, and others are still in discussions. Investigation of using these detectors to probe the RHNs have already been done in several existing literature. For example, Ref.~\cite{Kling:2018wct} has studied the RHNs from meson decays at the FASER.

On the other hand, the seesaw mechanism can be incorporated into ultra-violet complete models, with the $U(1)_{B-L}$ model being one of the simplest~\cite{Mohapatra:1980qe, Davidson:1978pm}. In such model, a $B-L$ gauge boson, $Z^\prime$ is also introduced, which can coupled to the quarks, and hence mesons. Therefore, $Z^\prime$ can be probed from meson decays at the FASER, as done in Ref.~\cite{Kling:2021fwx}. As the  $Z^\prime$ can decay into RHNs, so the $Z^\prime \rightarrow N \ N$ processes have also been used to probe the RHNs at the LHC~\cite{Deppisch:2019kvs}, as well as FCC-hh~\cite{Liu:2022kid}. Similar processes have also been considered in Ref.~\cite{Jodlowski:2020vhr, Jho:2020jfz} at FASER for sterile $\nu$-specific $U(1)_s$ model,  where
the $Z^\prime$ dominantly produced from bremsstrahlung process and scatterings off electrons. Since it has not discussed in details, so it becomes interesting to investigate the possibility to use $Z^\prime$ mediated RHNs from meson decays at the FASER for both the $U(1)_{B-L}$ and $U(1)_s$ Model. In this situation, the rates of the RHNs events should also depend on the coupling or the kinetic mixings, in addition to the $M_N$ as well as $|V_{lN}|^2$.
 
In this work, we study the sensitivity of FASER to the RHNs in the parameter space of ($M_N$,$|V_{lN}|^2$), via the  processes, meson $\rightarrow Z^\prime \rightarrow N \ N$ (including bremsstrahlung), at the high luminosity runs of the LHC~(HL-LHC), with $\mathcal{L} = 3000$ fb$^{-1}$. Different to the original $U(1)_{B-L}$ model, we also consider the case where the $Z^\prime$ is assumed to strongly interacted to the RHNs, with $g_X = 1$, while feebly to the SM particles, so-called the sterile $\nu$-specific $U(1)_s$ model.
These processes are sensitive to the RHNs with 0.1~GeV $< M_N <$ 10~GeV. We have taken the masses of the $Z^\prime$ from 0.5-2.5 GeV, or fixed $M_{Z^\prime} = 3 M_N$.
The gauge couplings or the kinetic mixings are taken at its maximal allowed value. The sensitivity of FACET is also studied, with no significant difference to the FASER obtained. 

The paper is organized as follows, in Section~\ref{sec:model}, we briefly review the $U(1)_{B-L}$ as well as $U(1)_s$ model. Next, w introduce the analyses of the signal events in Section~\ref{sec:events}. In Section~\ref{sec:results}, we show the resulting FASER sensitivity for different flavours. Finally, we conclude in Section~\ref{sec:end}.

\section{Model}
\label{sec:model}
\paragraph{$U(1)_{B-L}$ Model}
The relevant Lagrangian for the $Z^\prime$ and RHNs in the $U(1)_{B-L}$ model is~\cite{Bertuzzo:2018itn, Ballett:2018ynz}, 
\begin{align}
\label{L}
	\mathcal{L} \supset y_N  \bar{N^c} N \chi + y_D \bar{L}  N \tilde{H}+ g_{B-L} Y_{B-L} \bar{f} \gamma^\mu Z^\prime f + g_{B-L} Y_{B-L} \bar{N} \gamma^\mu Z^\prime N+ h.c.,
\end{align}
with $\tilde{H} = i \sigma^2 H^{*}$, $N$ the RHNs, and $f$ being the SM fermions plus the RHNs. $Y_{B-L}$ is the $U(1)_{B-L}$ quantum number. $\chi$ is $B-L$ scalar.

After the spontaneous symmetry breaking of the $U(1)_{B-L}$ and the SM, the completed mass matrix of the $N$ can be written as, 
\begin{align}
\label{matrix}
\mathcal{M}=\begin{pmatrix} 
0 & m_{D} \\ 
m_{D} &m_{R}
\end{pmatrix},
\end{align}
where 
\begin{align}
    m_D={\frac{y^{D}}{\sqrt{2}}}v,\qquad
    m_R=\sqrt{2}y^{N}x,
\end{align}
with $x$ being the vacuum expection value of the $B-L$ scalar.

In the seesaw limit where the masses of RHNs are much larger than the active neutrinos, i.e. $m_R\gg m_D$, their masses are approximately,
\begin{align}
    m_{\nu}\sim  -m_{D}m_{R}^{-1} m_{D}^{T}, \qquad
    M_{N}\sim m_{R},
\end{align}
which recovers the type-I seesaw.

The flavour and mass eigenstates of the light and heavy neutrinos are connected as  
\begin{align}
\label{Neutrino}
	\begin{pmatrix}
		\nu_L \\ \nu_R
	\end{pmatrix} = 
	\begin{pmatrix}
		V_{LL} & V_{LR} \\
		V_{RL} & V_{RR}
	\end{pmatrix}
	\begin{pmatrix}
		\nu \\ N
	\end{pmatrix},
\end{align} 
where $V_{LL} \approx U_{PMNS}$, $V_{RR}\sim 1$. So the states of the heavy neutrinos and RHNs are roughly the same. The active-sterile mixings $V_{LR}=V_{LR}\equiv V_{lN}$, control the interaction of the RHNs to the active neutrinos.

Now we move to the $Z^\prime$, which has been searched in various experiments. In our interested parameter space, with $M_{Z^\prime} \sim \mathcal{O}$(GeV), the current limits show $g_{B-L} \lesssim 2 \times 10^{-4}$~\cite{Ilten:2018crw}, as we take the maximal allowed value to get an optimistic sensitivity.

\begin{figure}
    \centering
    \includegraphics[width=0.35\textwidth]{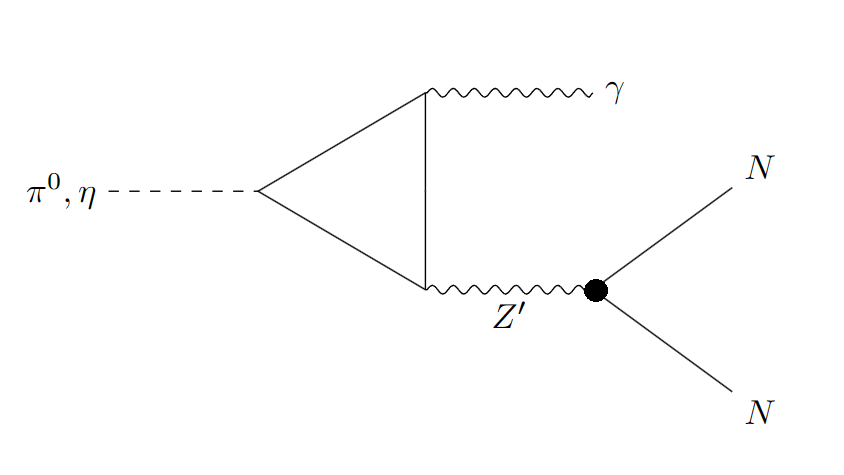}
    \caption{The Feynman diagram of the process  meson $\rightarrow Z'\rightarrow N \ N$}
    \label{fig:Feynman}
\end{figure}

For our main processes, meson $\rightarrow Z'\rightarrow N \ N$, the Feynman diagram is shown in Fig.~\ref{fig:Feynman}. The dominant mesons are the neutral pseudocscalar, $\pi^0$ and $\eta$ at the LHC, whereas the direct production of the $Z^\prime$ via bremsstrahlung is also considered.
Since the RHNs pairs are produced from the on-shell decays of the $Z^\prime$, the cross section of the signal processes can be estimated as 
\begin{align}
    \sigma_{S}=\sigma(\pi^0/\eta \rightarrow Z^\prime \gamma )\ \times BR(Z^\prime \rightarrow N \ N), 
\end{align}
where~\cite{Kling:2021fwx} 
\begin{align}
    \sigma(\pi^0/\eta \rightarrow Z^\prime \gamma )=\sigma_M \times 2 (g_{B-L}/e) ^2 \times BR(\pi^0/\eta \rightarrow \gamma \gamma) \times (1 - M_{Z^\prime}^2 / m_{\pi^0/ \eta}^2)^3,
\end{align}
where $\sigma_M$ is the cross section of mesons($\pi^0, \eta$) and $BR(Z^\prime \rightarrow N \ N) \approx 10 \%$ for each flavours of $N$ as taken from Ref.~\cite{Deppisch:2019kvs}.
$Z'$ can also be produced via proton bremsstrahlung, and the cross section of the signal processes can be estimated as 
\begin{align}
	\sigma_{S}=\sigma_B \times BR(Z^\prime \rightarrow N \ N),
\end{align}
where $\sigma_B$ is the cross section of $Z'$ produced from bremsstrahlung. Since we focus on the light RHNs, the production from Drell-Yan processes is negligible.

The $N$ subsequently decay dominantly via three body decays by an off-shell $W/Z$ boson. The expression of the decay width for $N$ dominantly coupled to $l_1$ flavour reads~\cite{Helo:2010cw}  
\begin{eqnarray}\label{lln-CC}
\Gamma(N\rightarrow l_1^{-}l_2^{+}\nu_{l_{2}} )&=& |V_{l_1 N}|^2
\frac{G_F^2}{192\pi^3} M_N^5 I_{1}(y_{l_1},y_{\nu_{l_{2}}}, y_{l_2})(1-\delta_{l_{1}l_{2}}), \\
\label{lln}
\Gamma(N\rightarrow \nu_{l_{1}}l_2^{-}l_2^{+} )&=& |V_{l_1 N}|^2
\frac{G_F^2}{96\pi^3} M_N^5
\left[\left(g^{l}_{L} g^{l}_{R}+ \delta_{l_{1}l_{2}}g^{l}_{R}\right) I_{2}(y_{\nu_{l_{1}}}, y_{l_{2}}, y_{l_{2}}) + \right. \\ 
&&\left.   + \left((g^{l}_{L})^{2} +(g^{l}_{R})^{2 }+ \delta_{l_{1}l_{2}} (1 +2 g^{l}_{L})\right) I_{1}(y_{\nu_{l_{1}}}, y_{l_{2}}, y_{l_{2}}) \right],
\\
\label{3n}
\sum_{l_{2}=e,\mu,\tau}\Gamma(N\rightarrow \nu_{l_{1}} \nu_{l_{2}} \bar{\nu}_{l_{2}})&=& |V_{l_1 N}|^2 
\frac{G_F^2}{96\pi^3} M_N^5  ,\\
\label{lP}
\Gamma(N\rightarrow l^{-}_{1} P^{+}) &=& |V_{l_1 N}|^2
\frac{G_F^2}{16 \pi}M_N^3 f_{P}^2 |V_{P}|^{2} F_P(y_{l_{1}},y_{P}),\\
\label{nuP}
\Gamma(N\rightarrow \nu_{l_{1}} P^0) &=& |V_{l_1 N}|^2 \frac{G_F^2}{64 \pi}M_N^3  f_{P}^2 
(1 - y^2_{P})^2,\\ 
\label{lV}
\Gamma(N\rightarrow l^{-}_{1} V^{+}) &=& |V_{l_1 N}|^2 
\frac{G_F^2}{16\pi}M_N^3 f_{V}^2 |V_{V}|^{2} F_V(y_{l_{1}},y_{V}),\\ 
\label{nuV}
\Gamma(N\rightarrow \nu_{l_{1}} V^0) &=&  |V_{l_1 N}|^2 \frac{G_F^2}{2 \pi}M_N^3 \ f_{V}^2\  \kappa_{V}^{2} (1 - y^2_{V})^2
(1 + 2 y^2_{V}),
\end{eqnarray} 
where kinematic functions $I_{(1,2)}(x,y,z)$, $F_{P,V}(x,y)$, leptonic couplings $g_{(L,R)}^{l}$ and meson decay constant $f_{(P,V)}$ are given in Ref.~\cite{Helo:2010cw}. And $y_i \equiv m_i/M_N$, with $m_i = m_P, m_V, m_l, m_q$.

\paragraph{Sterile $\nu$-specific $U(1)_s$ Model}
The relevant Lagrangian for the $Z^\prime$ and Dirac RHNs in the $U(1)_s$ model
is~\cite{Bertuzzo:2018itn,Jho:2020jfz},
\begin{align}
\label{Ls}
	\mathcal{L} \supset y_s  \bar{N} v_s \phi^\dagger + y_D \bar{L}  N \tilde{H}+ 
	g_X \epsilon \cos \theta_W Q \bar{f} \gamma^\mu Z^\prime f + g_{X} \bar{N} \gamma^\mu Z^\prime N+ h.c.,
\end{align}
with $\tilde{H} = i \sigma^2 H^{*}$, $N$ the RHNs, and $f$ being the SM fermions. $Y$ is the quantum number. $\chi$ is SM singlet scalar field.

The cross section of the signal processes now depend on the mixing $\epsilon$, 
\begin{align}
    \sigma(\pi^0/\eta \rightarrow Z^\prime \gamma )=\sigma_M \times 2 \epsilon ^2 \times BR(\pi^0/\eta \rightarrow \gamma \gamma) \times (1 - M_{Z^\prime}^2 / m_{\pi^0/ \eta}^2)^3,
\end{align}
and the current limits point $\epsilon \lesssim 6 \times 10^{-4}$ in our interested parameter space~\cite{Kling:2021fwx}.

In this model, we can have $g_X \sim 1 \gg \epsilon$, therefore the $BR(Z^\prime \rightarrow N \ N) \approx 100 \%$, if only one flavor of $N$ is coupled to the $Z^\prime$. Instead of decaying via an off-shell $W/Z$ boson at the $U(1)_{B-L}$ model, now the major decay channel is via an off-shell $Z^\prime$. Hence, the decay length of $N$ in this model can be much smaller than the $U(1)_{B-L}$ model for fixed $|V_{lN}|^2$. The decay channels via an off-shell $Z^\prime$ can be expressed as,
\begin{eqnarray}
\Gamma(N\rightarrow \nu_{l_{1}}l_2^{-}l_2^{+} )&=& |V_{l_1 N}|^2
\frac{G_X^2 \epsilon^2}{48\pi^3} M_N^5
[ I_{2}(y_{\nu_{l_{1}}}, y_{l_{2}}, y_{l_{2}}) 
+ 2 I_{1}(y_{\nu_{l_{1}}}, y_{l_{2}}, y_{l_{2}})]
,\\
\label{3ns}
\sum_{l_{2}=e,\mu,\tau}\Gamma(N\rightarrow \nu_{l_{1}} \nu_{l_{2}} \bar{\nu}_{l_{2}})&=& |V_{l_1 N}|^2 
\frac{G_X^2 \epsilon^2}{48\pi^3} M_N^5,\\
\label{nuPs}
\Gamma(N\rightarrow \nu_{l_{1}} P^0) &=& |V_{l_1 N}|^2 \frac{G_X^2 \epsilon^2}{32 \pi}M_N^3  f_{P}^2 
(1 - y^2_{P})^2,\\ 
\label{nuVs}
\Gamma(N\rightarrow \nu_{l_{1}} V^0) &=&  |V_{l_1 N}|^2 \frac{G_X^2 \epsilon^2}{\pi}M_N^3 \ f_{V}^2\  \kappa_{V}^{2} (1 - y^2_{V})^2
(1 + 2 y^2_{V}),
\end{eqnarray} 
where $G_X=g_X^2/(4 \sqrt{2} M_{Z^\prime}^2)$. 

\begin{figure}
    \centering
    \includegraphics[width=0.49\textwidth]{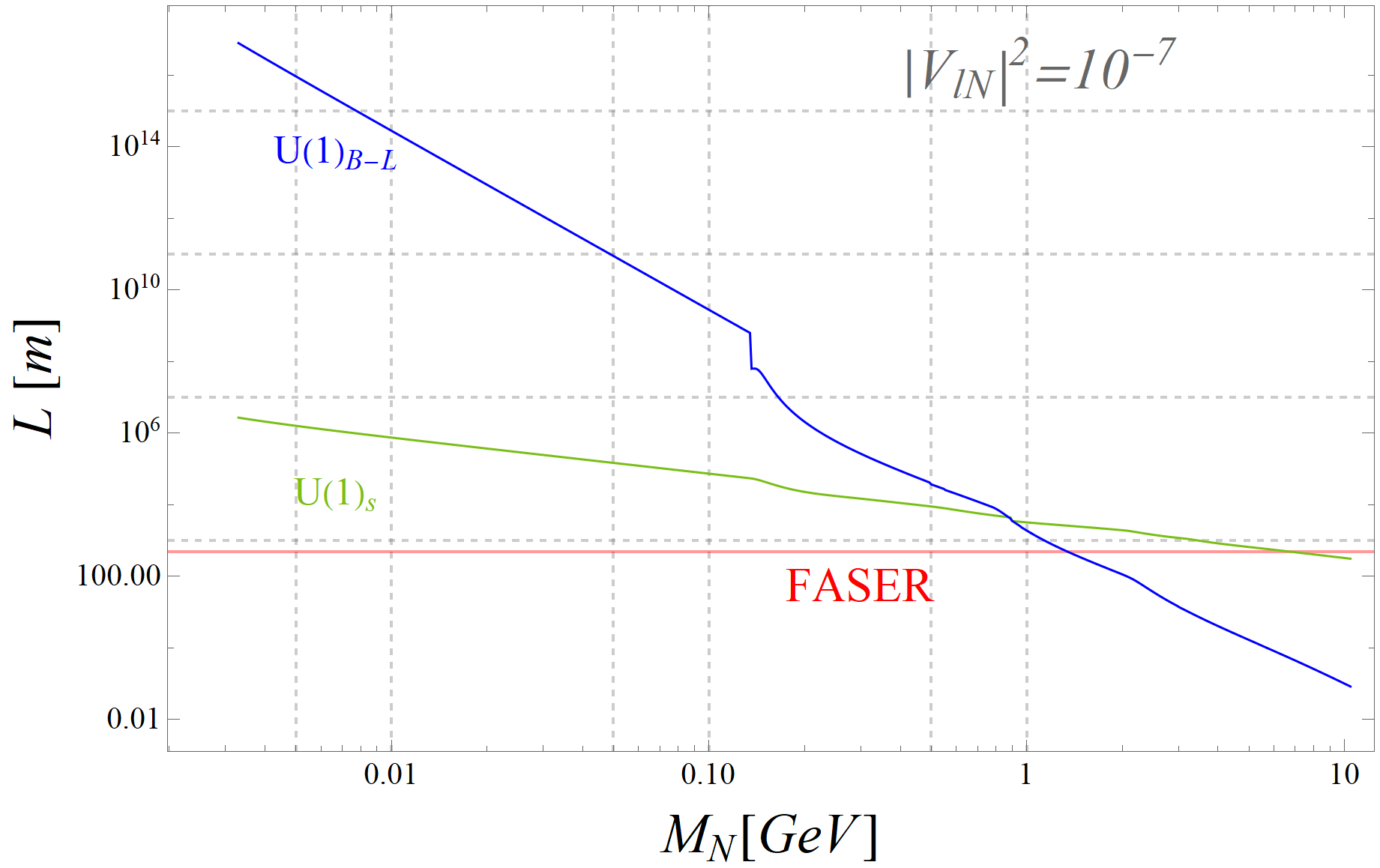}
    \includegraphics[width=0.49\textwidth]{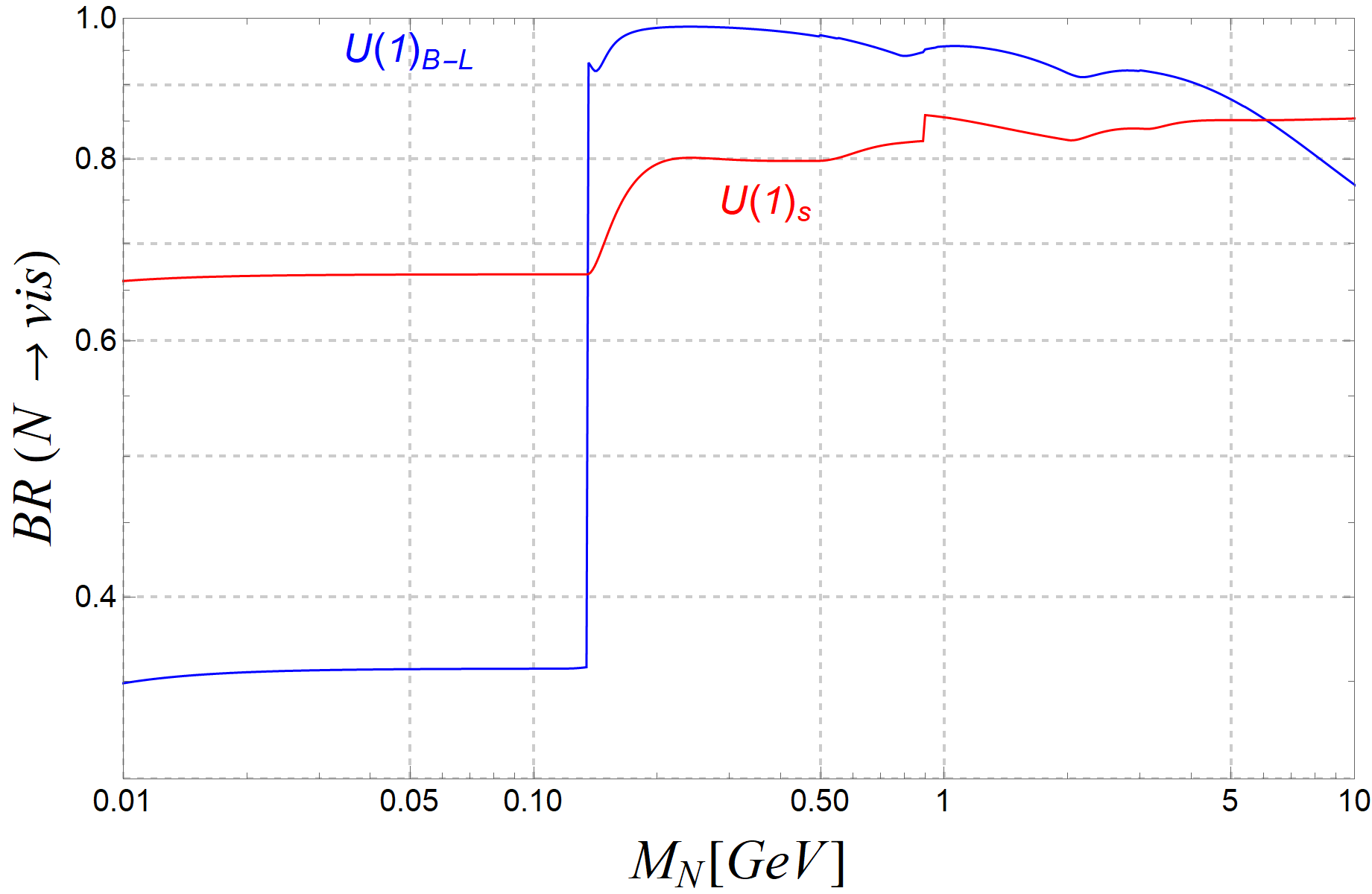}
    \caption{Left: The decay length of the $N$~ for the $U(1)_{B-L}$ and $U(1)_{s}$ model. Right: The branching ratio of the $N$ decays into visible final states for the two models. We fix $|V_{lN}|^2 = 10^{-7}$, where $l=e, \mu, \tau$ and the $N$ only couples to one flavor each, and $\epsilon = 6 \times 10^{-4}$, $g_X =$~1. }
    \label{fig:length}
\end{figure}

In Fig.~\ref{fig:length}~(left), we show the decay length of $N$ for the $U(1)_{B-L}$ and $U(1)_s$ model respectively, where we fix $\epsilon = 6 \times 10^{-4}$ at the current limit, $g_X =$ 1, and $|V_{lN}|^2 = 10^{-7}$. For the $U(1)_{B-L}$ model, in principle, the RHNs can also decay via an off-shell $Z^\prime$. Nevertheless, as we take $g_{B-L} \ll e$, the decay width is negligible comparing to the weak decays~\cite{Atre:2009rg}. Therefore, the decay width is only dependent on $|V_{l N}|^2$. For the $U(1)_s$ model,
we take $g_X = 1$, and $M_{Z^\prime} \ll M_Z$, hence $N$ decays via an off-shell $Z^\prime$ dominantly, and the decay width dependent on both $|V_{l N}|^2$ as well as $\epsilon$, $g_X$.
When $M_N \lesssim 1$ GeV, $L(N)$ in the $U(1)_{B-L}$ is larger than the $U(1)_s$ case, while it becomes the opposite when $M_N \gtrsim 1$ GeV.

The branching ratio of the $N$ decays into visible final states for the $U(1)_{B-L}$ and $U(1)_s$ models is shown in Fig.~\ref{fig:length}~(right). This is relevant to estimate the potential signal yield, as the experiments are more likely to observe visible final states, other than the $\nu \nu \nu$ final states. In the figure, we find that, once the $N$ is heavier than light hadrons, e.g., $\pi$ and $\eta$, $BR(N \rightarrow vis) \gtrsim 90\% $ for the $U(1)_{B-L}$ model, and above 80\% for the $U(1)_s$ model.

\section{Analyses}

\label{sec:events}
Now we estimate the number of signal events at the FASER,
which should be
\begin{align}
    N_{S}=\sigma(\pi^0/\eta \rightarrow Z^\prime \gamma ) \times BR(Z^\prime \rightarrow N \ N) \times BR^2(N \rightarrow vis) \times \mathcal{P} \times \mathcal{L},
\end{align}
where the $P$ is of possibility of $N$ decay inside the FASER volume, which is a function of momentum of the $N$, and $\theta$, the angle of the momentum and the beam line.
\begin{align}
    \mathcal{P}(p, \theta)=(e^{-(L-\Delta)/d}-e^{-L/d})\Theta(R-\tan\theta L)\approx \frac{\Delta}{d}e^{-L/d}\Theta(R-\theta L),
\end{align}
here $L$ is the distance of the FASER to the interaction point, $\Delta$ is the depth of the FASER volume, $d$ is the lab decay length of the $N$, and $R$ is the radius of the FASER. The lab decay length takes account the Lorentz boost factor, $d = c \tau \beta \gamma =  c \tau \times p / M_N$. 

Since $p, \theta$ of the $N$ is a distribution, so the number of signal events becomes,
\begin{align}
    N_S=\mathcal{L}\int dp d\theta \frac{d\sigma}{dpd\theta} \times \mathcal{P}(p,\theta)
\label{eq:Ns}
\end{align}

The FASER detector is going to be operated in two phase~\cite{FASER:2018eoc,Feng:2017uoz}
,
\begin{align}
\text{\textbf{FASER1}}:L = 480~\text{m}, \Delta = 1.5~\text{m},  R = 10~\text{cm},  \mathcal{L} = \text{150~fb}^{-1},\nonumber\\
\text{\textbf{FASER2}}:L =480~ \text{m}, \Delta = 10~\text{m},  R = 1~\text{m},  \mathcal{L} = \text{3~ab}^{-1}.
\end{align}
Since the volume of FASER1 is too small, in the following calculation, we focus on the FASER2 setup. There is also a trigger requirement on the RHNs, such that $p > $ 100 GeV.

We also consider the FACET detector~\cite{Ovchynnikov:2022its}, with
\begin{align}
\text{\textbf{FACET}}:L = 101~\text{m}, \Delta = 11~\text{m},  R = 0.18-0.5~\text{m},  \mathcal{L} = \text{3~ab}^{-1}.
\end{align}

\begin{figure}[h]
    \centering
    \includegraphics[scale=0.45]{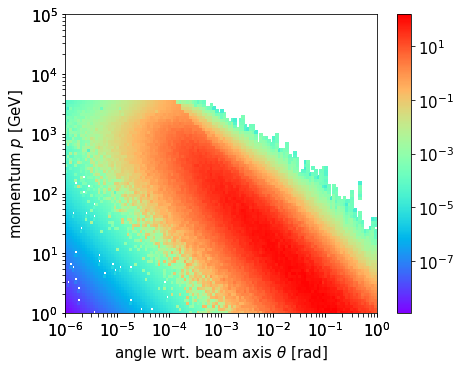}
    \caption{The spectrum of the momentum, $p$ and angle to the beam axis $\theta$ for the $Z^\prime$, with $M_{Z^\prime} =$ 0.1 GeV and $g_{B-L} = 2 \times 10^{-4}$, generated by using {\tt FORESEE} package~\cite{Kling:2021fwx}.}
    \label{fig:N_Spectrum}
\end{figure}

In order to estimate the number of signal events via the aforementioned methods, we use the FORESEE python package in Ref.~\cite{Kling:2021fwx}. The kinematic distribution of various mesons in the one forward hemisphere have already been stored in the FORESEE package. Once we provide the formulas for the production of the $Z^\prime$ from mesons as well as bremsstrahlung, the package can output the differential spectra ($d\sigma(M \rightarrow Z^\prime)/dp_{Z'} dcos\theta_{Z'}$) of the $Z^\prime$, where $p_{Z'}$ is the 3-momentum and $\theta_{Z'}$ is the polar angle.
The kinematic distribution of the $Z^\prime$ is shown in Fig.~\ref{fig:N_Spectrum}. As $Z^\prime$ decays into two identical $N$, so $p_N \approx p_{Z^\prime}/2$, and $\theta_N \approx \theta_{Z^\prime}$ since $p_N \gg M_N$ in the forward direction. Therefore, the distribution of the $p, \theta$ for the $N$ should be very similar to the ones for the $Z^\prime$. From Fig.~\ref{fig:N_Spectrum}, we can see that, since $Z^\prime$ is mainly produced via the decays of mesons, therefore their transverse momentum $p_T \sim \Lambda_{QCD}$, that is why most of the particles are distributed at the forward direction. 
For the FASER2 location where $\theta \sim 2 \times 10^{-3}$, and the trigger requirement $p > 100$~GeV, there are still hundreds of events, and we use Eq.~\ref{eq:Ns} to estimate the number of signal events at the FASER2.

\section{Results and discussions}

\label{sec:results}
In this section, we show the expected sensitivity of FASER2 to the right-handed neutrinos in the selected parameter space. We take 3000 ${\text{fb}}^{-1}$ integrated luminosity for FASER2. Since the $N$ can decay into pairs of visible particles, their vertex can be reconstructed at the FASER2. And the detectors are shields from hundreds of meters from the IP of the LHC, so the background in this case can be negligible, hence we only require $N_S > 3$  to define the sensitivity at 95\% confidence level.

Overall, the number of the signal events depends on four free parameters, ($M_N$, $|V_{lN}|^2$, $M_{Z^\prime}$, $g_{B-L}/ \epsilon$). In order to probe the RHNs, i.e. to obtain sensitivity in the ($M_N$, $|V_{lN}|^2$) plane, we need to fix the $M_{Z^\prime}$ and $g_{B-L}/ \epsilon$. As mentioned in the text, we always fix the coupling or the mixing $g_{B-L}/ \epsilon$ at the maximally allowed values by the current limits. For $M_{Z^\prime}$, we begin with fixed values of 0.5, 1.0, 1.4, 2.0 and 5.0 GeV, and followed by fixed ratios of $M_{Z^\prime} = 3 M_N~$.

\begin{figure}[htbp]
    \centering
    \includegraphics[width=0.49\textwidth]{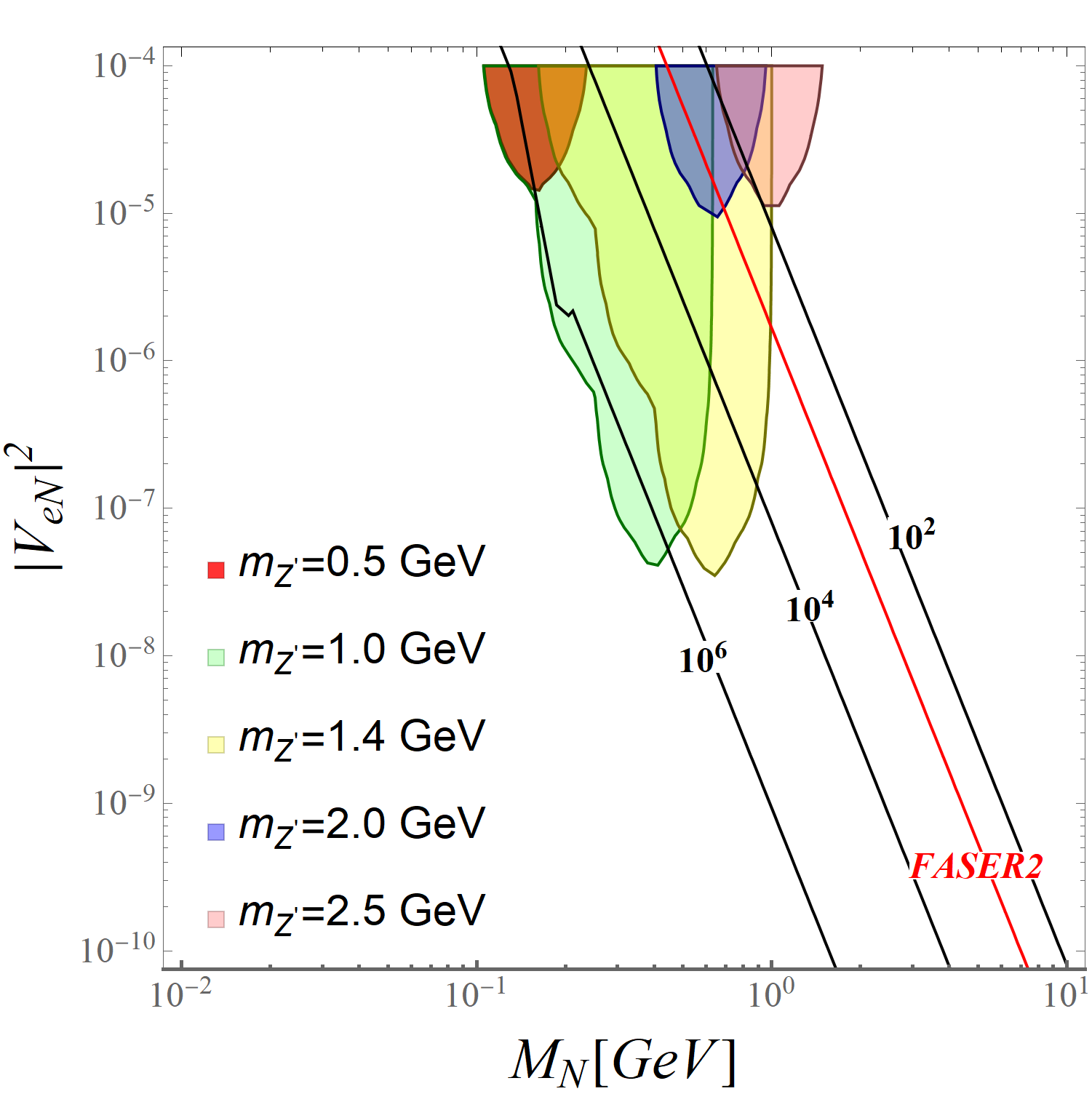}
     \includegraphics[width=0.49\textwidth]{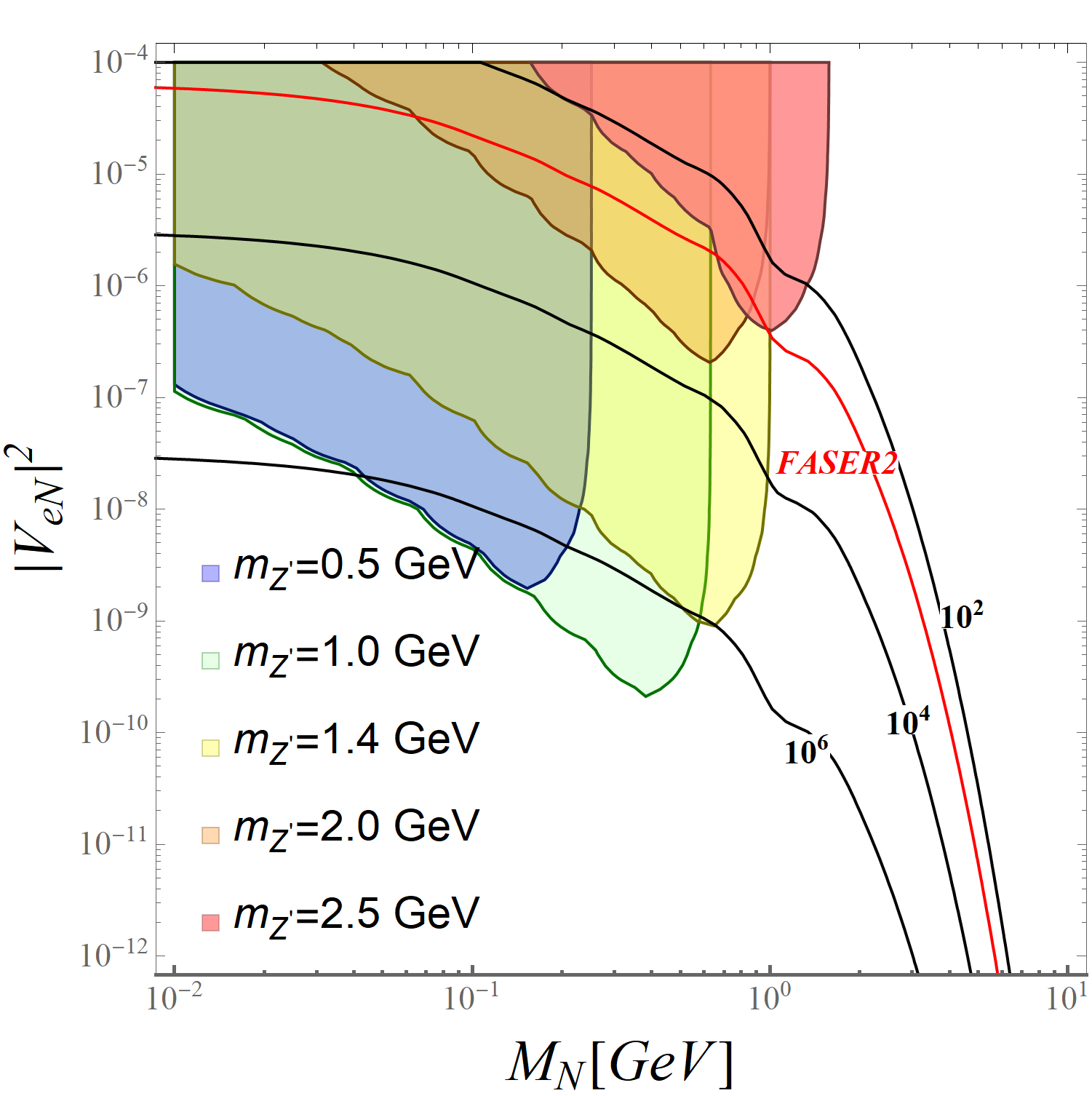}
    \caption{Left: The sensitivity on the ($M_N$, $|V_{eN}|^2$) plane of FASER2, for the $U(1)_{B-L}$ model. We fixed $M_{Z^\prime} = 0.5, 1.0, 1.4, 2.0, 2.5$~GeV, as indicated by red, green, yellow, blue and pink bands, respectively. The curves represent the fixed values of the proper decay length of $N$, $L(N) = 10^{2,4,6}$, and 480 meters which is the distance from the FASER2 to the LHC's IP are overlaid for comparison. The gauge coupling is fixed at $g_{B-L} = 2 \times 10^{-4}$. 
    Right: the same but for the $U(1)_{s}$ model, and the kinetic mixing is fixed at $\epsilon = 6 \times 10^{-4}$.}
    \label{fig:fixmzp}
\end{figure}

In Fig.~\ref{fig:fixmzp}, we show the sensitivity on the ($M_N$, $|V_{eN}|^2$) plane, for both the $U(1)_{B-L}$~(left) and $U(1)_{s}$~(right) models. The solid curves represent the fixed values of the proper decay length of $N$, $L(N) = 10^{2,4,6}$, and 480 meters which is the distance from the FASER2 to the LHC's IP are overlaid for comparison. Since the $N$ is largely boosted at the forward direction, the lab decay length can even be much larger, e.g. by $10^3$ times.
For the $U(1)_{B-L}$ model, the best sensitivity is obtained $|V_{eN}|^2\sim 10^{-7}$ where $M_N \approx 0.3$~GeV, when $M_{Z^\prime} = 1~(1.4)$~GeV. In such scenario, the $Z^\prime$ is heavy enough, to let the $N$ to have decay length close to the FASER2's distance to the IP, and not so heavy so the $Z^\prime$ can not be produced by meson decays and bremsstrahlung. When it comes to the $U(1)_{s}$ model, benefited from the decay length of the $N$ much closer to the FASER2 for fixed $|V_{eN}|^2$ and larger decay branching ratio from $Z^\prime$, now we can get almost three more magnitude better sensitivity, as we approach $|V_{eN}|^2\sim 10^{-10}$ where $M_N \approx 0.5$~GeV, when $M_{Z^\prime}=$~1~GeV.

\begin{figure}[h]
     \centering
     \includegraphics[scale=0.5]{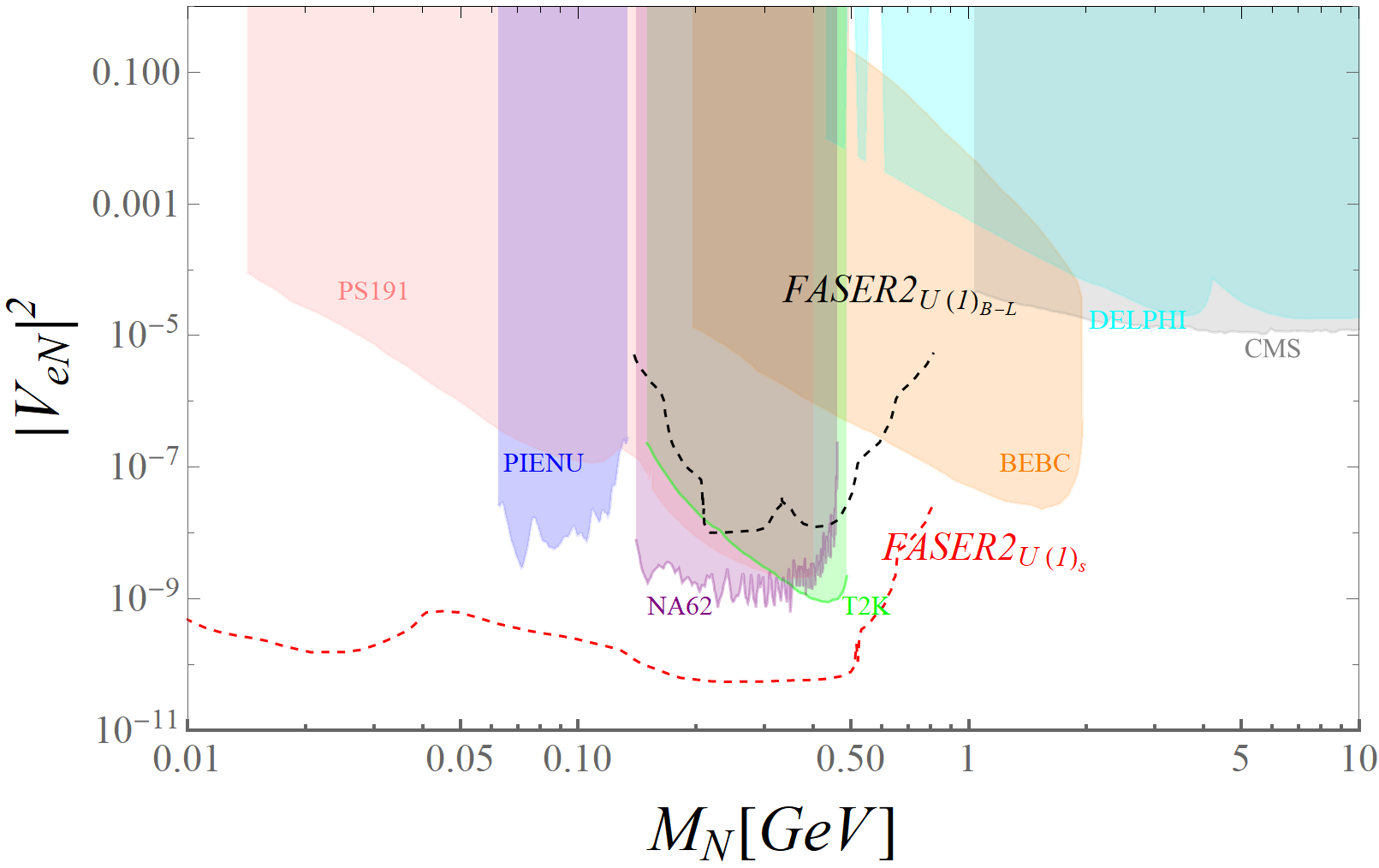}
     \caption{The sensitivity on the ($M_N, |V_{l N}|^2$) of the FASER2 for the $U(1)_{B-L}$~(dashed black) and $U(1)_s$~(dashed red) model, when $l=e$. Current limits from NA62~\cite{NA62:2020mcv}, T2K~\cite{T2K:2019jwa},  PIENU~\cite{PIENU:2017wbj}, DELPHI~\cite{DELPHI:1996qcc}, CMS~\cite{CMS:2018iaf},BEBC~\cite{Barouki:2022bkt} and PS191~\cite{Bernardi:1987ek} are overlaid for comparison.}
     \label{fig:eN}
 \end{figure}

\begin{figure}[h]
     \centering
     \includegraphics[scale=0.5]{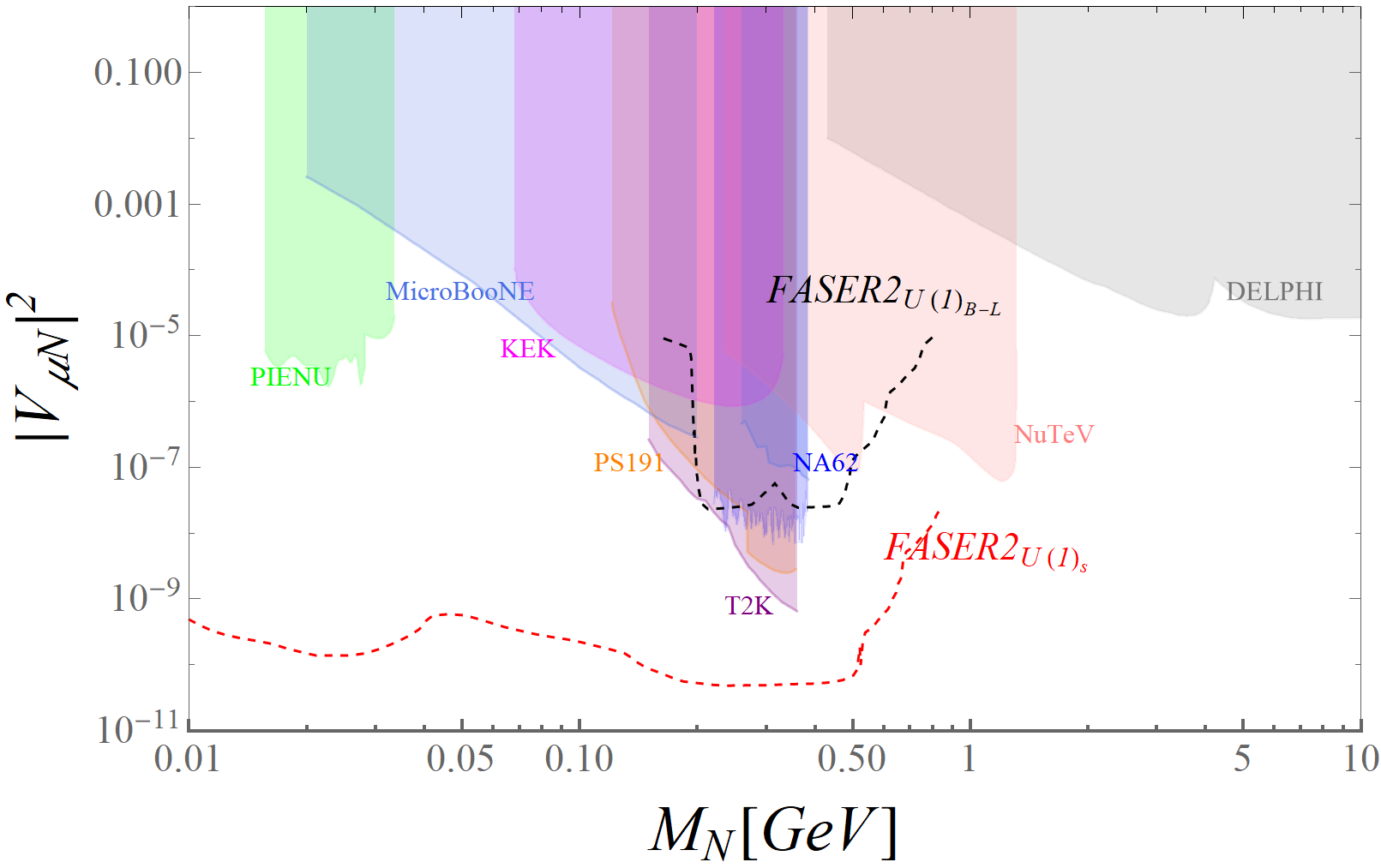}
     \caption{The same to Fig.~\ref{fig:eN}, but for $l=\mu$. Current limits from NA62~\cite{NA62:2020mcv}, T2K~\cite{T2K:2019jwa},  PIENU~\cite{PIENU:2017wbj}, DELPHI~\cite{DELPHI:1996qcc}, PS191~\cite{Bernardi:1987ek}, KEK~\cite{E949:2014gsn}, NuTeV~\cite{NuTeV:1999kej} and MicroBooNE~\cite{MicroBooNE:2019izn} are overlaid for comparison.}
     \label{fig:muN}
 \end{figure}

\begin{figure}[h]
     \centering
     \includegraphics[scale=0.5]{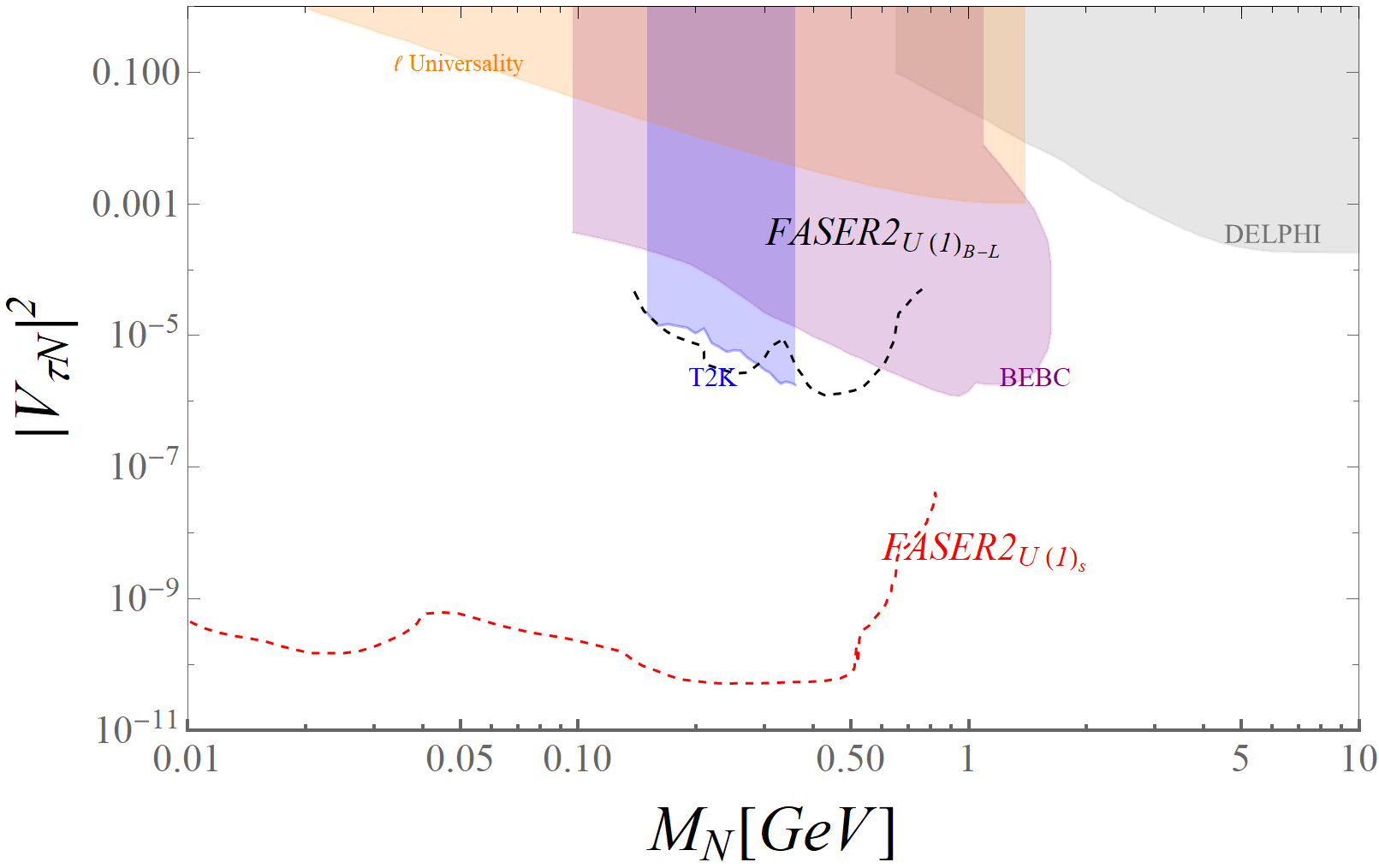}
     \caption{The same to Fig.~\ref{fig:eN}, but for $l=\tau$. Current limits from T2K~\cite{T2K:2019jwa}, BEBC~\cite{Barouki:2022bkt}, DELPHI~\cite{DELPHI:1996qcc} and lepton universality test~\cite{Cvetic:2017gkt}.}
     \label{fig:tauN}
 \end{figure}

Now we move to the scenarios where we fix $M_{Z^\prime} = 3 M_N$. The results are shown in Fig.~\ref{fig:eN}, ~\ref{fig:muN}, and~\ref{fig:tauN}, where the results are compared with the current limits from different existing experiments for the RHNs to be dominated coupled to $e$, $\mu$, $\tau$ leptons, respectively. In our interested parameter space, for the electron flavour, the most powerful exisiting constraints come from the beam dump experiments NA62~\cite{NA62:2020mcv} as well as T2K~\cite{T2K:2019jwa}, mainly from the process $K\rightarrow l N$, $l = e, \mu$, hence sensitive to $M_N \approx m_K \approx 0.5~$GeV. Both of the detectors are put $\mathcal{O}(10^2)$ meters away from the beam, similar to the FASER. Other beam dump experiments, PIENU~\cite{PIENU:2017wbj}, BEBC~\cite{Barouki:2022bkt} and PS191~\cite{Bernardi:1987ek} also searched for the RHNs, whereas their constraints are not competitive as the aforementioned ones. Nevertheless, the PS191 and PIENU searched the $\pi \rightarrow l N$, and BEBC searched the $\tau \rightarrow X X N$, thus
can be sensitive to $M_N \lesssim 0.2$~GeV and $M_N \gtrsim 0.5$~GeV
which is not probed by NA62 and T2K. In addition, the RHNs have also been searched via the direct production at electron-positron colliders, $e^+ e^- \rightarrow Z \rightarrow N \nu$ at the DELPHI experiment at the LEP~\cite{DELPHI:1996qcc}. When it comes to the muon flavour, the different masses of the muon and electron, translate the constraints to the lower $M_N$. In addition, KEK~\cite{E949:2014gsn}, NuTeV~\cite{NuTeV:1999kej} and MicroBooNE~\cite{MicroBooNE:2019izn} also searched for RHNs with the muon flavour.  For the tau flavour, since the tau lepton is hard to detect, so only a few experiments including T2K, BEBC, DELPHI as well as the lepton universality test~\cite{Cvetic:2017gkt} are available, with much looser constraints.

Unlike the case for fixed $M_{Z^\prime}$, now the kinematics threshold requiring $Z^\prime$ to decay into two $N$s is spontaneously satisfied, hence the reach to the mass of the $N$s extends to wider regions. Despite of the different reconstruction efficiencies for the leptons at the FASER2 which is unknown yet, the sensitivity from the FASER2 should not depend on the flavour of the RHNs.  For the $U(1)_{B-L}$ model, the sensitivity of the FASER2 reaches the lowest $|V_{lN}|^2 \approx 10^{-8}$ when $M_N \sim 0.3$ GeV, so the $Z^\prime$ is roughly resonantly produced by the decay of the $\eta$ meson. The sensitivity on $|V_{lN}|^2$ becomes worse drastically to $|V_{lN}|^2 \approx 10^{-5}$, when we move to the lower and larger $M_N$, since either the decay length is too large, or the production cross section is too small. 

For the $U(1)_{B-L}$ model, comparing to the existing experiments on the electron flavour of the RHN, FASER2 roughly has the same sensitivity to the ones from the NA62 and T2K. When we consider RHNs heavier than $m_K$, the sensitivity of the FASER2 has exceed the current best limits from the BEBC. The situation is similar on the muon flavour, only now current best limits for RHNs heavier than $m_K$ is from the NuTeV, and also has been exceed by the FASER2 via $Z^\prime$ decays. When it comes to the $\tau$ flavour, since the decay channel $N \rightarrow \tau^- X X$ is forbidden in most of our interested parameter space, so the decay length of the $N$ becomes much larger, so the sensitivity can only reach larger $V_{lN}^2$, which is close to the current limits as well.

As mentioned, in the $U(1)_s$ model, the FASER2 should have much better sensitivity on the active-sterile mixings $|V_{lN}|^2$. Indicated by the red dashed lines in Fig.~\ref{fig:eN}, ~\ref{fig:muN}, and~\ref{fig:tauN}, the FASER2 can now probe the mixing as low as $|V_{lN}|^2\sim 10^{-10}$, which is roughly two magnitude better than the ones in the  $U(1)_{B-L}$ model. This is better than the current limits for all three flavours of the RHNs. And now the FASER2 is sensitive to $M_N$ as low as 0.01 GeV, as the decay length for such light $N$ is no longer depend on the mass of $N$, but the kinetic mixing $\epsilon$ in this model.
Hence, in this case, the FASER2 has shown positive potential to reveal the origin of the neutrino mass problem, i.e. probing the RHNs.
However, for the existing beam dump experiments, in order to get the right sensitivity, recast should be done to consider the different decay length of $N$ in the appearance of the $U(1)_s$ $Z^\prime$.


\section{Conclusion}

As the seesaw mechanism explained the neutrino oscillation experiment, the important goal of the study of neutrino mass is to find the right-handed neutrino. According to the seesaw mechanism, $m_\nu \sim |V_{lN}|^{2}M_N$~\cite{Deppisch:2019kvs}, hence at the current limits for sub GeV RHNs, they can be regarded as long-lived particles.
So the goal of our work is to look for RHNs and to estimate the sensitivity of the far detector FASER2 to the RHNs.

In this work we explored the sensitivity of FASER2 detector to the RHNs in meson $\rightarrow Z'\rightarrow N \ N$ channel. Both the $U(1)_{B-L}$ and sterile $\nu$-specific $U(1)_s$ model are considered. In the former model, the $Z^\prime$ couples to all the fermion with the same gauge coupling, which we fixed at $g_{B-L} = 2 \times 10^{-4}$. While for the latter one, the $Z^\prime$ coupling to the RHNs can be sufficiently higher than the ones for SM fermions, $g_X \sim 1 \gg \epsilon = 6 \times 10^{-4}$. Hence, in the $U(1)_s$ model, the $Z^\prime$ almost only decay into RHNs, and the decay of the RHNs is dominated by a off-shell $Z^\prime$. For similar couplings and $|V_{lN}|^2$, the branching ratio of $Z^\prime$ to the RHNs is higher, and the decay length of the RHNs is shorter and closer to the distance of the FASER2 to the IP, if $M_N \lesssim 1$ GeV.

Two scenarios of the masses of $Z^\prime$ are discussed. In the first one, the masses are fixed at $M_{Z^\prime} = $ 0.5, 1.0, 1.4, 2.0 and 5.0 GeV. In the second one, the ratio between the masses of the RHNs and $Z^\prime$ is fixed at $M_{Z^\prime} = 3 M_N$. In both scenarios, we estimate the number of signal events, after considering the decay length of the RHNs, and the geometrical information of the volume of the FASER2. With no background assumuption, sensitivity to the $(M_N, |V_{lN}|^2)$ is obtained by asking $N_S > 3$.

We then compare our results with existing limits. Depending on the dominated couplings to the leptons of the RHNs, the current limits can be classified into three categories, the limits on ${\lvert V_{eN}\rvert}^2$, ${\lvert V_{\mu N}\rvert}^2$ and ${\lvert V_{\tau N}\rvert}^2$, respectively.
Fig~\ref{fig:eN} gives  the current constraints of electron mixing  ${\lvert V_{eN}\rvert}^2$ which has been obtained experimentally. The experiments include PIENU~\cite{PIENU:2017wbj}, T2K~\cite{T2K:2019jwa}, NA62~\cite{NA62:2020mcv},BEBC~\cite{Barouki:2022bkt} and DELPHI~\cite{DELPHI:1996qcc}. Fig~\ref{fig:muN}  indicates the current limits on muon mixing ${\lvert V_{\mu N}\rvert}^2$ from NA62~\cite{NA62:2020mcv}, T2K~\cite{T2K:2019jwa},  PIENU~\cite{PIENU:2017wbj}, DELPHI~\cite{DELPHI:1996qcc}, PS191~\cite{Bernardi:1987ek}, KEK~\cite{E949:2014gsn}, NuTeV~\cite{NuTeV:1999kej} and MicroBooNE~\cite{MicroBooNE:2019izn}. For the $U(1)_{B-L}$ model, as can be seen from the figures, our results reach to $|V_{lN}|^2 \sim 10^{-8}$, are comparable to the current limits overall. When the mass of $N$ is less than 0.5~GeV~($m_K$), the results obtained by the current experiment are better. However, when the mass of $N$ is more than 0.5~GeV~($m_K$) and less than about 0.6 GeV, our results are better than the current limits from BEBC/NuTeV. Fig~\ref{fig:tauN}  indicates the current limits on tau mixing ${\lvert V_{\tau N}\rvert}^2$ from T2K~\cite{T2K:2019jwa}, BEBC~\cite{Barouki:2022bkt} , DELPHI~\cite{DELPHI:1996qcc} and lepton universality test~\cite{Cvetic:2017gkt}. Now the current limits as well as the limits in this paper are rather poor, since the decay into $\tau$ leptons final states is kinematically forbidden, and $N$ are too long-lived to be detected by the FASER detector. When we move to the $U(1)_s$ model, the gain in larger $N$ production from $Z^\prime$ decay, and closer decay length to the FASER2's volume lead to about two magnitude better sensitivity on $|V_{lN}|^2$. For much lighter $N$ with $M_N < 0.1$~GeV, since the decay length is controlled by the kinetic mixing $\epsilon$ instead of $M_N$, the sensitivity curves become roughly constant at $|V_{lN}|^2 \sim 10^{-10}$, where all the current limits can not reach. 

We have shown that in certain scenarios, FASER can yield better sensitivity than the current limits. However, the current limits are performed by sterile neutrinos only models, such that the RHNs are directly produced via the mixing from the weak decays and etc. 
Proper comparison between different detectors can be made, if all the limits can be recast to the same model. However, even though the recast of the production can be feasible, the kinematic and detector efficiencies still require full Monte-carlo simulation, which we leave for future works.

We have obtained similar sensitivity from FACET, comparing to the FASER. Since the $Z^\prime$ is dominantly produced via light mesons $\pi^{0}$ and $\eta$, $Z^\prime$ and hence the RHNs are more likely to distribute in a very forward direction, where the FASER located. FACET, although placed closer to the transverse direction, its coverage in solid angle is several times larger. Hence, in this study, the overall effects lead to similar sensitivity for both FACET and FASER. In a recent work, the processes that the RHNs are produced via the decays of $D$ and $B$ mesons are considered~\cite{Ovchynnikov:2022its}. As the RHNs are now produced directly from massive mesons, they are distributed closer to the transverse direction, hence the FACET turns out to be more sensitive.

\label{sec:end}

\acknowledgments

WL is supported by National Natural Science Foundation of China (Grant No.12205153), and the 2021 Jiangsu Shuangchuang (Mass Innovation and Entrepreneurship) Talent Program (JSSCBS20210213). HS is supported by the National Natural Science Foundation of China (Grant No.12075043). We thank Arindam Das for useful discussions.

\bibliography {submit}
\end{document}